\documentclass[
 reprint, amsmath,amssymb, aps, prx, superscriptaddress, twocolumn, longbibliography, notitlepage, nofootinbib, 10pt]{revtex4-2}

\usepackage{graphicx}
\usepackage{dcolumn}
\usepackage{bm}
\usepackage{soul}
\usepackage{centernot}
\usepackage{textgreek}
\usepackage{siunitx}
\usepackage[dvipsnames]{xcolor}

\usepackage[breaklinks=true,colorlinks,citecolor=color1,linkcolor=color1,urlcolor=color1]{hyperref}
\usepackage{color}
\definecolor{color1}{rgb}{0,0,1}
\definecolor{color3}{rgb}{0.85,0,0.5}
\newcommand{\sfref}[2]{\textcolor{color1}{\hyperref[#1]{Fig.$\,$\ref{#1}(#2)}}}

\newcommand{\aref}[1]{\textcolor{color1}{\hyperref[#1]{App.$\,$\ref{#1}}}}

\begin{document}

\preprint{APS/123-QED}

\title{Nonlocal Josephson coupling and current redistribution in a large-area graphene multiterminal Josephson junction}

\author{Luojia Zhang}
\email{lz282@cornell.edu}
\affiliation{Department of Physics, Cornell University, Ithaca, NY, USA}

\author{Kushagra Aggarwal}
\affiliation{School of Applied and Engineering Physics, Cornell University, Ithaca, NY, 14853, USA}

\author{Jay D. Sau}
\affiliation{Condensed Matter Theory Center and Joint Quantum Institute, Department of Physics,
University of Maryland, College Park, Maryland 20742-4111, USA}

\author{Ethan G. Arnault}
\affiliation{Department of Electrical Engineering and Computer Science, Syracuse University, Syracuse, NY 13244, USA}
\affiliation{Institute for Quantum and Information Sciences, Syracuse University, Syracuse, NY 13244, USA}

\author{Kenji Watanabe}
\affiliation{Research Center for Electronic and Optical Materials, National Institute for Materials Science, 1-1 Namiki, Tsukuba 305-0044, Japan}

\author{Takashi Taniguchi}
\affiliation{Research Center for Materials Nanoarchitectonics, National Institute for Materials Science, 1-1 Namiki, Tsukuba 305-0044, Japan}

\author{Valla Fatemi}
\email{vf82@cornell.edu}
\affiliation{School of Applied and Engineering Physics, Cornell University, Ithaca, NY, 14853, USA}

\date{\today}

\begin{abstract}
Multiterminal Josephson junctions (MTJJs) extend conventional two-terminal Josephson physics to devices defined by multiple superconducting phases, where rich phenomena have been predicted.
However, in conventional MTJJs, superconducting transport is often dominated by local coupling between neighboring leads, obscuring physics that goes beyond simple connectivity.
Here, we realize a multiterminal Josephson network in a graphene Hall bar geometry with micron-scale dimensions, enabling substantial nonlocal coupling between distant terminals.
Our key observation is a pronounced negative differential longitudinal resistance (NDLR) in the four-probe response near switching currents, while no such feature appears in two-terminal measurements.
We reproduce the NDLR response with a network containing the interplay between two distinct forms of nonlocality: nonlocal Josephson coupling, which determines the switching sequence of the superconducting network, and nonlocal normal-state conductance, which redistributes current after different regions switch.
Our results highlight the potential of the emergent NDLR for characterizing superconductor--semiconductor hybrid systems and for nonlinear circuits.
Further, this Josephson Hall bar geometry provides a platform for exploring the interplay between mesoscopic superconductivity and many-body ground states in graphene-based heterostructures.
\end{abstract}

\maketitle

\section{Introduction}
Multiterminal Josephson junctions (MTJJs) provide a versatile platform for exploring emergent quantum phenomena as well as for developing quantum coherent devices~\cite{Riwar2016,Aasen2016Majorana,Draelos2019,Pankratova2020,Gupta2023GateTunableDiode,Yokoyama2015,Arnault2021,TopoTranscond,Arnault2022Multiplet}.
In an $n$-terminal MTJJ, the $n$ superconducting leads impose $n-1$ independent phase differences, allowing both the supercurrent and the Andreev bound-state spectra to be controlled within an ($n-1$)-dimensional phase space, rather than by a single phase bias as in a conventional junction~\cite{Riwar2016,Draelos2019,Pankratova2020,Gupta2023GateTunableDiode,Yokoyama2015,Arnault2021,Arnault2022Multiplet}.
This enhanced tunability gives rise to a wealth of novel phenomena, including nontrivial current–phase relations~\cite{Gupta2023GateTunableDiode,Pankratova2020}, phase-coherent transport across multiple terminals~\cite{QuartetHuang,QuartetTomo,Multiplet,Melin2023PiQuartet,Arnault2021,Arnault2022Multiplet}, the prospect of single-fermion manipulation~\cite{vanHeck2014}, and the simulation of topological band structures~\cite{Riwar2016,TopoTranscond,Yokoyama2015,Houzet2019MajoranaWeyl,Aasen2016Majorana,Strambini2016OmegaSQUIPT,Meyer2017Chern}.

\begin{figure}[htb]
\centering
\includegraphics[]{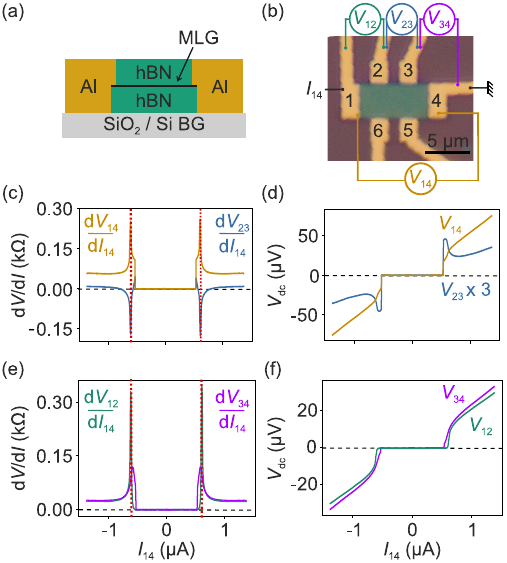}
\caption{(a) Device structure of the Josephson Hall bar.
(b) Optical image of the Josephson Hall bar with the measurement configuration.
(c),(e) Differential resistance measured under a source--drain bias $I_{14}$, showing $dV_{23}/dI_{14}$ and $dV_{14}/dI_{14}$ in (c), and $dV_{12}/dI_{14}$ and $dV_{34}/dI_{14}$ in (e). Red dashed lines indicate the position of the NDLR.
(d),(f) Corresponding dc voltages $V_{23}$ and $V_{14}$ in (d), and $V_{34}$ and $V_{12}$ in (f), as functions of $I_{14}$.}\label{fig:fig1}
\end{figure}

Most of the previously studied MTJJs with weak links made of semiconductor two-dimensional electron gases, diffusive normal metals, or encapsulated graphene were highly compact, with distances between superconducting electrodes ranging from tens to hundreds of nanometers~\cite{Graziano2020ThreeTerminalJJ,Pankratova2020,Gupta2023GateTunableDiode,Graziano2022}.
These length scales are typically comparable to or smaller than the mean free path in the normal region to ensure that quasiparticles can traverse the device without excessive suppression of supercurrent.
In this regime, the device physics is dominated by correlations between nearest-neighbor electrodes.
Consequently, the supercurrent and Andreev bound state spectrum are largely governed by nearest-neighbor phase differences, while nonlocal interactions between other pairs of terminals are weak. 
The dominance of nearest-neighbor coupling constrains access to a variety of interesting physics from nonlocal effects, such as crossed Andreev reflection and elastic cotunneling~\cite{Deutscher2000,Russo2005,CaddenZimansky2006,CaddenZimansky2007,Feng2025}, phase dragging~\cite{Omelyanchouk2000,Amin2001,Draelos2019}, and magnetic focusing of electrons~\cite{Chen2016,Taychatanapat2013}.

In this work, we present a distinct MTJJ structure based on a Hall bar geometry, in which the normal region extends for several microns in both planar directions.
We refer to the device as a Josephson Hall bar.
The monolayer graphene normal region, with its long mean free path, can support substantial supercurrent in such large devices~\cite{Borzenets2016Ballistic,Calado2015Ballistic}.
We show that this enlarged structure supports nonlocal coupling between all pairs of terminals, consistent with near-ballistic transport.
We distinguish two forms of nonlocality in this network.
The first is nonlocal Josephson coupling, referring to coherent superconducting coupling between distant terminals mediated by Andreev trajectories.
The second is nonlocal normal coupling, referring to the normal-state conductance network that redistributes dissipative current after switching.
This gives rise to a pronounced negative differential longitudinal resistance (NDLR) in the four-terminal measurement configuration in the vicinity of switching currents.

This NDLR results from a two-stage process. 
First, nonlocal Josephson couplings determine which supercurrent branches switch into the resistive state, providing a necessary nonlinearity.
Second, the resulting resistive network redistributes current through the nonlocal normal conductance channels, which can produce a reduction in the measured four-probe voltage. 
These results establish the Josephson Hall bar as a versatile platform for studying nonlocal superconducting transport and emergent collective switching phenomena. 

\section{Results}
The Josephson Hall bar device consists of a six-terminal monolayer graphene Hall bar encapsulated by hexagonal boron nitride (hBN) and contacted using the edge-contact geometry. 
The Hall bar is 3 $\mu$m wide, with a source–drain separation of 6 $\mu$m.
Details of the device fabrication are given in the SI. An optical image of the Josephson Hall bar device is shown in \sfref{fig:fig1}{b}. The device was measured in a dilution refrigerator.
Unless otherwise stated, all measurements were performed at a mixing-chamber temperature of 10~mK.

We first characterize the transport by measuring the current--voltage characteristics under dc current bias.
The evolution of the measured voltage reveals multiple switching features occurring at distinct current biases, indicating that different node pairs undergo the superconducting-to-normal transition sequentially rather than simultaneously.
In a four-probe configuration, where the internal voltage $V_{23}$ is measured, we observe a pronounced NDLR that emerges shortly after the switching current (\sfref{fig:fig1}{c}).
Direct current--voltage characterization confirms the NDLR as a clear back-bending of the four-terminal transfer curve between $V_{23}$ and $I_{14}$ within the NDLR regime (\sfref{fig:fig1}{d}).
\sfref{fig:fig1}{e,f} show the ac and dc measurements characterizing the switching of the regions near the source ($V_{12}$) and drain ($V_{34}$). 
The onset of NDLR in $dV_{23}/dI_{14}$ coincides with the switching of the source-side voltage difference into a finite-voltage state, evidenced by the peak in $dV_{12}/dI_{14}$, and occurs simultaneously with the peak in the two-terminal response $dV_{14}/dI_{14}$.
The slight difference between the source-side and drain-side switching currents likely arises from device inhomogeneity.
As we will show, the NDLR is inconsistent with a nearest-neighbor-only connectivity model, in which each junction switches independently and transport is governed only by the superconducting phase differences of neighboring leads.

\begin{figure}[h]
\centering
\includegraphics[]{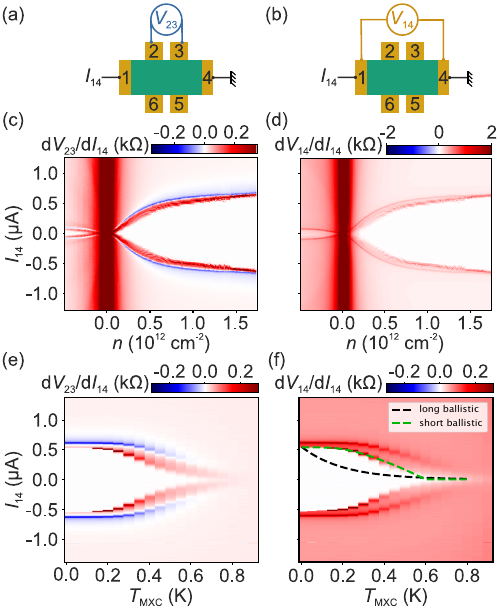}
\caption{(a), (b) Measurement configuration for $dV_{23}/dI_{14}$ and $dV_{14}/dI_{14}$, respectively. (c), (d) $dV_{23}/dI_{14}$ and $dV_{14}/dI_{14}$ as a function of $I_{14}$ and $n$. (e), (f) $dV_{23}/dI_{14}$ and $dV_{14}/dI_{14}$ as a function of $I_{14}$ and $T_\text{MXC}$, at a fixed density of $1\times10^{12}\,\mathrm{cm}^{-2}$. Panel (f) shows the expected critical current as a function of temperature for long (black dashed line) and short ballistic (green dashed line) junctions.
}\label{fig:fig2}
\end{figure}

\begin{figure*}[htb]
\centering
\includegraphics[]{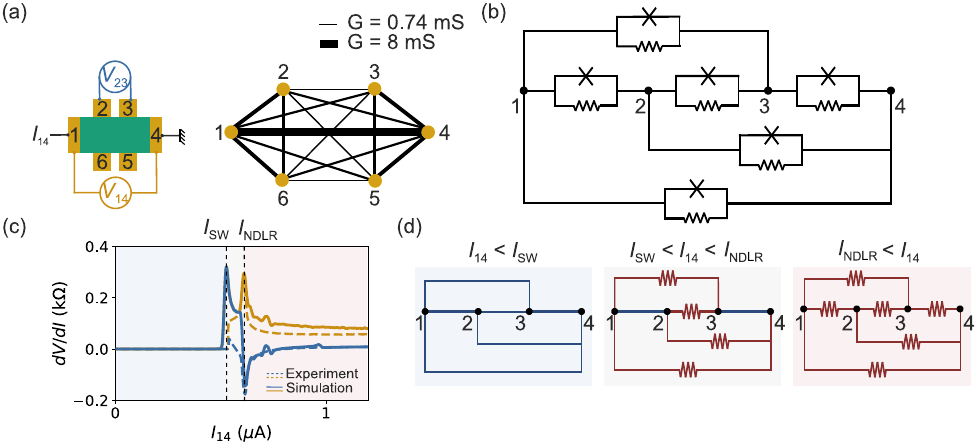}
\caption{(a) Device measurement scheme (left) and extracted conductance matrix (right), where the thickness of the lines between nodes $i$ and $j$ is proportional to the conductance $G_{ij}$. (b) Effective RSJ network model for the Josephson Hall bar device under dc $I_{14}$ current bias. (c) Comparison of experimental data with simulation. $\mathrm{I_\text{sw}}$ indicates the current at which the node pair 2-3 switches to a finite voltage state, and $\mathrm{I_{NDLR}}$ is the current at which the NDLR is maximized. (d) Equivalent circuits at different $I_{14}$ regimes, color-coded as in (c). Blue indicates that the edge is superconducting, and red indicates that it is in a finite-voltage state. The critical-current parameters used in (c) are: $I_{c,12}=I_{c,34}=250\,\mathrm{nA}$, $I_{c,23}=12\,\mathrm{nA}$, $I_{c,13}=I_{c,24}=190\,\mathrm{nA}$, $I_{c,14}=178\,\mathrm{nA}$.}
\label{fig:fig3}
\end{figure*}

We performed a study of carrier-density and temperature-dependence of the NDLR.
The measurement configuration is presented in \sfref{fig:fig2}{a,b}.
A current bias $I_{14}$ was applied, and voltages $V_{23}$ and $V_{14}$ were measured.
The gate dependence of the transport provides further insight into the origin of the effect (\sfref{fig:fig2}{c,d}).
On the electron-doped side, supercurrent emerges above a carrier density of approximately \(0.3\times10^{12}\,\mathrm{cm^{-2}}\), coinciding with the onset of NDLR. On the hole-doped side, the NDLR is likewise observed, although the overall supercurrent is weaker due to the presence of \(p\text{--}n\) barriers in the graphene near the contacts \cite{Borzenets2016Ballistic,Calado2015Ballistic,Jang2024Engineering}.
Across the full gate range, the NDLR evolves smoothly with carrier density and is observed only in the internal four-probe voltage.
This suggests that the NDLR is a reproducible device response of the Josephson Hall bar, rather than an artifact arising from the measurement setup or a result of fine-tuning.

The temperature dependence further supports this interpretation (\sfref{fig:fig2}{e,f}).
The \(I\)--\(V\) characteristics exhibit no measurable hysteresis, indicating that the NDLR is unlikely to arise from thermal runaway or switching instabilities~\cite{Courtois2008,Shcherbakova2015}.
Moreover, the effect persists up to \SI{0.6}{\kelvin}, exceeding half of the superconducting critical temperature \(T_c\) of the leads (\SI{1}{\kelvin}).
In \sfref{fig:fig2}{f}, we compare the measured temperature dependence of the multiterminal network switching current with the expected behavior of short- and long-ballistic junctions.
The observed dependence is most consistent with a junction in or near the short-ballistic regime, in agreement with the long carrier mean free path and high Fermi velocity in graphene.
This interpretation further suggests that the induced minigap is comparable in magnitude to the parent superconducting gap. 
For completeness, we also consider the possibility that some transport channels lie in the long-ballistic regime.
In that case, the characteristic minigap may be estimated as $E_{\mathrm{gap}}=\hbar v_Fl/d^2\approx 0.64\ \mathrm{K}$, where $v_F,l,d$ denote the graphene Fermi velocity, carrier mean free path (\SI{\sim 3}{\micro\meter}, inferred from the Drude model), and characteristic device dimension, respectively~\cite{Beenakker2004AndreevBilliards}. 
We emphasize, however, that the commonly used analytical expressions are generally derived for two-terminal junctions. 
In a MTJJ, nonlocal coupling between terminals can produce a more complicated dependence of the critical current on junction dimensions and temperature~\cite{Bobkov2026,Polevoy2025}.

To establish the generality of the observed phenomena, we fabricated six different devices, including examples with monolayer and bilayer graphene as normal regions, and examples with niobium and aluminum as superconducting contacts (see the SI).
Despite device-to-device variations in superconducting contact materials, graphene thickness, gate design, and device geometry, all devices exhibit the characteristic NDLR behavior, demonstrating the robustness across the tested devices.
This insensitivity to device details suggests that the effect is not set by a specific material parameter or contact property, but instead originates from a network-level mechanism rather than a finely tuned property of one device.
We also note that previous work has pointed out that negative differential resistance could arise from a Josephson junction coupled to a cavity~\cite{PEDERSEN2014178}.
However, this mechanism is unlikely here, as we have fabricated devices with different on-chip environments (e.g., different substrates and on-chip wiring conductivities).

\subsubsection{Origin of the NDLR}

By measuring multiple configurations of current source--drain and voltage probes, we built a conductance-network model that reveals substantial nonlocal (off-diagonal) terminal conductances. The measurements were performed at a fixed density of $1\times10^{12}\,\mathrm{cm}^{-2}$.
The extracted conductance matrix is visualized in \sfref{fig:fig3}{a}, where the thickness of the black solid lines represents the magnitude of the corresponding conductance.
Details of the extraction of this matrix are given in the SI.
We find that the strongest conduction channel corresponds to nonlocal coupling between nodes 1 and 4, consistent with the larger number of transport modes supported by the wide contacts 1 and 4.
In contrast, the direct 2--3 (and similarly 5--6) paths exhibit the weakest conductance. This is consistent with a predominantly ballistic transport picture, in which electrons injected from node 2 have a low probability of reaching node 3.

We connect the experimentally extracted network to the observed NDLR using resistively shunted junction (RSJ) simulations.
The junction capacitance here is neglected due to the small geometric capacitance in the device.
Exploiting the approximate mirror symmetry about the 1--4 axis, as reflected by the nearly symmetric extracted conductance matrix (see SI), the basic physics of the full six-node network can be reduced to an effective four-node model.
Between each pair of nodes $i$ and $j$ we include an effective RSJ model with critical current $I_{c,ij}$ shunted by a normal resistor $R_{ij}$. 
The experimentally extracted conductance matrix determines the normal resistive channels $R_{ij}$, while the Josephson critical currents $I_{c,ij}$ describe the superconducting coupling between terminals.
The current on each link is
\begin{equation}
I_{ij}
= I_{c,ij}\,\sin(\phi_i-\phi_j)
+ \frac{\Phi_0}{2\pi R_{ij}}\,
(\dot{\phi}_i-\dot{\phi}_j),
\end{equation}
where $\phi_i$ denotes the superconducting phase of node $i$ and $\Phi_0=h/2e$ is the superconducting flux quantum.
The first term includes the nonlocal Josephson coupling, whereas the second includes current flow through the nonlocal normal conductance network after switching. 
Imposing current conservation at each node yields a set of coupled nonlinear differential equations, which we solved numerically.
The dc voltage between nodes $(i,j)$ is $V_{ij}=\frac{\Phi_0}{2\pi}\langle\dot{\phi}_i-\dot{\phi}_j\rangle_t$.
The critical currents in each link $I_{c,ij}$ were chosen to best reproduce the experimental data (see SI).
The simulations qualitatively reproduce the essential experimental features, including the emergence of NDLR near the switching current, as shown in \sfref{fig:fig3}{c}. 
 
Qualitatively, the NDLR can be understood as a redistribution of current between local and nonlocal conduction pathways as different regions of the network undergo switching.
The equivalent circuits under current bias in each regime are illustrated in \sfref{fig:fig3}{d}.
For bias currents below the first switching current $I_\text{sw}$, all phase differences remain locked and the device is fully superconducting.
Just above $I_\text{sw}$, the central junction along the 2--3 link switches to a finite-voltage state, while the junctions near the left and right ends remain superconducting. As the current bias is further increased, the 1--2 and 3--4 junctions also switch to the resistive state. This additional switching redistributes the current away from the 2--3 edge, reducing the current flowing through that branch. Consequently, a decrease in $V_{23}$ occurs concurrently with switching near the source and drain nodes, giving rise to the observed NDLR. 

Furthermore, we support the origin of the NDLR by modifying the Josephson coupling between the leads in two different device geometries (details in the SI). 
In one approach, the graphene channel is fully segmented by two intermediate superconducting contacts, effectively forming three Josephson junctions in series.
This device did not exhibit NDLR. 
In another approach, we increased the distance of the source and drain electrodes from the interior region to \SI{4}{\micro\meter}, so that no supercurrent flows between the source or drain and the interior region. In both cases, no NDLR is observed.
These results support the interpretation that the observed NDLR originates from nonlocal coupling and switching-induced current redistribution within an extended Josephson network, rather than from a local junction instability.

\subsubsection{Flux dependence}
\begin{figure*}[htb]
\centering
\includegraphics[]{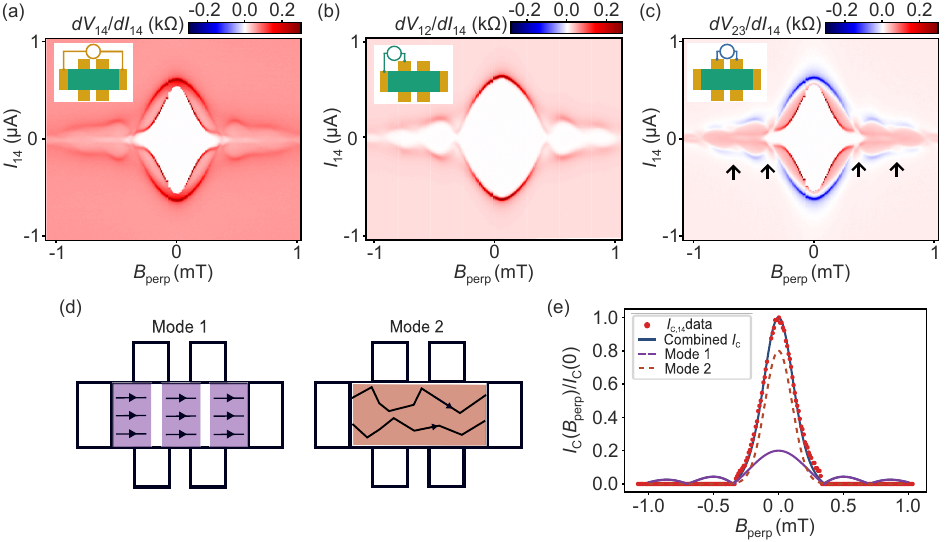}
\caption{\label{fig:epsart}(a)--(c) Differential resistance as a function of perpendicular magnetic field $B_{\mathrm{perp}}$ and current bias $I_{14}$, measured across the global $1\!-\!4$ response, the source-side $1\!-\!2$ region, and the central $2\!-\!3$ region, respectively. Arrows in (c) mark magnetic-field values where the NDLR is suppressed. 
(d) Schematic interpretation of the two magnetic-field-dependent superconducting contributions. Mode 1 represents a contact-adjacent contribution with a relatively well-defined interference area and sinc-like field dependence, while mode 2 represents an additional spatially extended contribution involving phase averaging over Andreev trajectories through the Hall-bar interior.
(e) Normalized critical current extracted from the $1\!-\!4$ response in (a), together with a two-component fit consisting of the mode-1 sinc-like contribution and the mode-2 Gaussian-like contribution. 
}\label{fig:fig5}
\end{figure*}

Finally, we show the evolution of this physics with magnetic field.
Unlike gate voltage and temperature, which primarily affect all junctions proportionally, magnetic flux affects phase-coherent loops.
Remarkably, this device supports well-defined Fraunhofer patterns.
No Fraunhofer periodicity corresponding to the full lithographic area of the device is observed, indicating that the device does not behave as a single two-terminal weak link from source to drain.
Distinct regions instead exhibit markedly different flux-dependence behaviors.
In particular, the two-terminal source--drain measurement, as shown in \sfref{fig:fig5}{a}, reveals a zero-resistance contour that coincides with the zero-resistance region observed in the four-probe internal voltage, indicating that global switching is constrained by the central region of the device.
Superimposed on this, $dV_{12}/dI_{14}$ in \sfref{fig:fig5}{b} displays a broader envelope of differential-resistance peaks associated with switching near the source and drain electrodes.
Notably, in \sfref{fig:fig5}{c}, the onset and evolution of the negative differential longitudinal resistance closely follow the switching of the region near the source electrode.
These observations support the coexistence of multiple spatially distinct current-carrying modes with different effective areas, consistent with transport governed by a distributed and strongly coupled Josephson network rather than a single coherent junction.
We further note that the NDLR vanishes at the Fraunhofer node, where the supercurrent of the region near the source and drain is suppressed.
This further supports that supercurrent-carrying channels between multiple pairs of terminals are required for the NDLR to appear, and it is consistent with our control experiment with longer normal regions near the source and drain that intrinsically do not support supercurrent (see the SI).

The flux dependence of the interior region is consistent with two distinct magnetic-interference contributions. 
\sfref{fig:fig5}{d} shows the spatial patterns for these two contributions.
Mode 1 ($I_{c,1}(B)$) exhibits a conventional sinc-like Fraunhofer dependence with a periodicity corresponding to an effective area of 5.9 $\mu\mathrm{m}^2$, matching that observed independently in the $1\!-\!2$ measurement.
We associate mode 1 with short-range superconducting coupling between neighboring contacts, such as the $1\!-\!2$, $2\!-\!3$, and $3\!-\!4$ regions.

Mode 2 ($I_{c,2}(B)$), in contrast, may be characterized by a monotonic, approximately Gaussian suppression with magnetic field. Such a response can arise from ensemble averaging over long Andreev trajectories that traverse the interior of the junction and enclose a distribution of magnetic areas, causing their accumulated phases to progressively dephase with increasing field~\cite{Chiodi2012}.
As shown in \sfref{fig:fig5}{e}, the critical-current data are well described by a weighted sum of the two contributions, $w_1I_{c,1}(B)+w_2I_{c,2}(B)$ with $w_1=0.2$ and $w_2=0.8$, indicating that the inner/global switching boundary is dominated by the Gaussian-like mode 2, with a smaller but finite contribution from mode 1.

The sequential switching of the device then explains why the long-range contribution is absent from the later $1\!-\!2$ switching boundary. When the $2\!-\!3$ and global $1\!-\!4$ responses first become resistive, the corresponding phase differences begin to evolve in time, so neither the local $2\!-\!3$ contribution nor the long-range mode 2 can continue to support a stationary dc supercurrent. In contrast, $V_{12}$ and $V_{34}$ remain zero, indicating that the $1\!-\!2$ and $3\!-\!4$ regions remain locally phase locked. Their subsequent switching is therefore governed predominantly by the surviving local mode-1 coupling, giving rise to the conventional sinc-like Fraunhofer envelope observed in $V_{12}$.

\section{Discussion}

The physics observed in the Josephson Hall bar geometry should be generic to any superconductor-semiconductor materials platform, such as those based on the InAs 2DEG family \cite{Lutchyn2018,Graziano2020ThreeTerminalJJ,Pankratova2020,Gupta2023GateTunableDiode,Graziano2022} and Ge 2DHG family \cite{Tosato2023,Luethi2023,Lakic2025}. 
In particular, the physics of the NDLR may be useful for benchmarking electronic superconducting transport properties at zero magnetic field. 
However, a missing piece to this end is a complete, quantitative model for the physics of the device, which will need to account for the distinct energy-phase relations exhibited by short, intermediate length, and long SNS channels~\cite{Golubov2004,Beenakker1991,Dubos2001}, normal scattering mechanisms that can mix transport trajectories that link different sets of electrodes, and nonequilibrium quasiparticle dynamics within the large number of channels involved~\cite{CaddenZimansky2006,Tinkham1972,Martinis2009,Clarke1972}. 
The same extended geometry may also provide a favorable addition to devices investigating crossed Andreev reflection and elastic cotunneling.

The NDLR gives the Josephson Hall bar a nonlinear dc response that is continuously tunable \textit{in situ}. 
The four-probe differential resistance becomes negative over a finite bias window, reaching values of order of \SI{-0.2}{\kilo\ohm}, and its magnitude evolves smoothly with carrier density and current bias. 
The associated voltage scale is small (\SI{\sim8}{\micro\volt} at \SI{\sim0.5}{\micro\ampere}) and dissipated power is on the order of \SI{10}{\pico\watt}, consistent with operation at millikelvin temperatures. 
In other devices, negative differential resistance underpins circuit functions such as amplification and mixing, which motivates asking whether the present effect could play a useful role as a low impedance nonlinear element.
Measurements of the high-frequency response, together with geometries in which the negative differential resistance couples to well-defined ports, can be used to assess this possibility.

More broadly, the terminal count is itself a resource. 
Some proposals have no analogue below five terminals, such as the recently predicted quantization of transconductance in five-terminal contact geometry, coupled to a symmetric electromagnetic environment~\cite{Mertiri2026}, which complements the topological transconductance quantization predicted for four-terminal Josephson junctions~\cite{TopoTranscond}.
The six-terminal Hall bar with independent current biases provides sufficient connectivity. 
Accessing the proposed physics would require engineering the on-chip electromagnetic environment so that quantum phase fluctuations are important, which can be done with standard on-chip superconducting resonator nanofabrication methods~\cite{Blais2021}.

Beyond its nonlinear functionality, the large junction geometry of the Josephson Hall bar provides a platform for exploring superconducting proximity effects in graphene systems hosting exotic ground states, including spin--orbit--engineered graphene~\cite{grapheneCPS,Island2019SOCBLG,Zhang2023Enhanced,Li2024Bernal} and rhombohedral graphene~\cite{zhou2021superconductivity,han2025chiral,zhou2021quartermetal,lu2024fractional,Han2024Pentalayer}.
In these systems, correlated and/or topological phases emerge from the bulk electronic structure and require spatially extended, electrostatically controlled regions to be probed reliably.
In conventional Josephson junctions, the short source--drain separation leads to a normal region that is strongly influenced by contact-material-induced doping and boundary effects, potentially obscuring bulk-like material properties.
Moreover, the limited junction length complicates the integration of high-quality van der Waals top gates, which are often necessary to achieve independent displacement-field and density tunability.
In contrast, the present device features a micron-scale normal region, enabling superconducting correlations to couple to a larger and more uniform graphene area that reduces the influence of the contacts and provides access to a larger interior region.
The open Hall bar architecture is also naturally compatible with spatially varied gating schemes, facilitating the definition of local mesoscopic regions in the interior, such as point contacts and quantum dots.
The Josephson Hall bar may therefore serve as a building block for investigating the interplay of superconductivity, topology, and many-body effects in graphene-based correlated materials and mesoscopic structures.

In conclusion, we demonstrated that a large graphene Hall bar can operate as a multiterminal Josephson device, forming a spatially extended and strongly coupled Josephson network. The micron-scale geometry enables pronounced nonlocal coupling between spatially separated terminals, such that transport is governed by the collective dynamics of a network containing long-range links rather than a simple nearest-neighbor series network.
This nonlocality manifests as current-controlled negative differential longitudinal resistance that closely follows switching near the source electrode and appears prominently in four-probe internal measurements, consistent with redistribution of current following Josephson switching.
The magnetic-field dependence further reveals multiple interference modes with distinct effective areas, consistent with spatially nonuniform and nonlocal supercurrent flow.

\begin{acknowledgments}
L.Z. acknowledges support from the David W. Meehl Fellowship, established in memory of James R. Meehl, and administered by the Cornell College of Engineering.
We gratefully thank the Aref and Manon Lahham Faculty Fellowship for support of L.Z. and this project.
K. A. and V. F. acknowledge support from the Kavli KICK grant LS-2025-GR-0232-3098.
J.D.S. acknowledges support from the Joint Quantum Institute and the Laboratory for Physical Sciences through their support of the Condensed Matter Theory Center at the University of Maryland.
K.W. and T.T. acknowledge support from CREST (JPMJCR24A5), JST, and the World Premier International Research Center Initiative (WPI), MEXT, Japan.
Research was partly sponsored by the Army Research Office and was accomplished under Grant Number W911NF-22-1-0053. The views and conclusions contained in this document are those of the authors and should not be interpreted as representing the official policies, either expressed or implied, of the Army Research Office or the U.S. Government. The U.S. Government is authorized to reproduce and distribute reprints for Government purposes notwithstanding any copyright notation herein.
This work was performed in part at the Cornell NanoScale Facility, a member of the National Nanotechnology Coordinated Infrastructure (NNCI), which is supported by the National Science Foundation (NSF) (Grant No. NNCI-2025233).

V.F. acknowledges helpful discussions with Anton Akhmerov and Christoph Strunk.
V.F. and L.Z. acknowledge helpful discussions with Leonid Levitov and Regis Melin.

\textbf{Author Contributions:}
L.Z. fabricated the device and collected the data with the help of K.A. and under the supervision of V.F.; J.D.S. proposed the theoretical model. E.A. contributed to the RSJ simulation script. L.Z., K.A., and V.F. wrote the manuscript with input from all authors.

\textbf{AI Disclosure:} We used Claude (Haiku 4.5) to help optimize the RSJ numerical model to run more efficiently and to assist with plotting. We also used ChatGPT (OpenAI) to assist with editing and polishing the manuscript text.
\end{acknowledgments}

\appendix

\bibliography{references}

\end{document}